\documentclass[12pt]{article}

\usepackage{graphicx}
\usepackage{newtxtext}
\usepackage{newtxmath}
\usepackage{natbib}
\usepackage{hyperref}
\hypersetup{
    colorlinks = true,
    urlcolor   = blue,
    citecolor  = black,
}

\newcommand{\RomanNumeralCaps}[1]
\linenumbers

\title{On Satisfying the Kutta Condition in Unsteady Thin Aerofoil Theory}

\author{Kiran Ramesh \footnote{kiran.ramesh@glasgow.ac.uk, School of
  Engineering, University of Glasgow, Glasgow, G12 8QQ, UK}}

\def\Put(#1,#2)#3{\leavevmode\makebox(0,0){\put(#1,#2){#3}}}

\newcommand\numberthis{\addtocounter{equation}{1}\tag{\theequation}}

\begin{document}

\maketitle 

\begin{abstract}

Unsteady thin-aerofoil theory is a low-order method for solving potential-flow aerodynamics on a camber-line undergoing arbitrary motion. In this method, a Kutta condition must be applied at the trailing edge to uniquely specify the net circulation about the aerofoil. This article provides a critical discussion on applying the Kutta condition in unsteady flows, and introduces an improved method of doing so in unsteady thin-aerofoil theory. Specifically, the shed wake at any discrete time step is represented by a continuous distribution of vorticity derived from the exact Wagner solution rather than by a point vortex or regularized vortex blob. Results in the article illustrate the effects of this improvement for cases of step change in angle of attack (Wagner problem), harmonic heaving motion (Theodorsen problem), and a pitch-ramp-hold manoeuvre. Exact analytical solutions and CFD simulations of the incompressible Euler equations are used for verification. The new approach is seen to satisfy the Kutta condition at all reduced frequencies, with velocities being finite and pressure difference going to zero at the trailing edge. It improves unsteady thin-aerofoil theory in terms of theoretical rigour, computational cost and numerical accuracy.  

\end{abstract}

\iffalse 
\section*{Nomenclature}
{\renewcommand\arraystretch{1.0}
\noindent\begin{longtable*}{@{}l @{\quad=\quad} l@{}}
$A_0, A_1 ... A_N$ & Fourier coefficients in bound vorticity distribution\\
$c$ & aerofoil chord\\
$U_{ref}$ & reference velocity for nondimensionalization\\
$t$ & time \\
$t^*=t U_{ref}/c$ & nondimensional time\\
$\Delta t$ & time step \\
$C_l$ & lift coefficient\\
$\alpha$ & aerofoil pitch angle\\
$h$ & aerofoil plunge displacement\\
$u$ & aerofoil horizontal velocity\\
$U_X, U_Z$ & freestream velocities\\
$W$ & downwash\\
$x$ & chordwise coordinate\\
$\theta$ & transformed chordwise coordinate\\
$\gamma$ & bound vorticity distribution on airfoil\\
$\gamma_w$ & wake vorticity distribution\\
$\Gamma_B$ & aerofoil bound circulation\\
$\Gamma_w$ & last-shed vortex circulation\\
$W_0$ & change in downwash at the $3/4$-chord location\\
$\Phi$ & Wagner function \\
$\Psi$ & Kussner function \\
$S$ & Sears function \\
$R$, $Q$ & functions used in analytical solutions of unsteady flows
\end{longtable*}}

\fi

\section{Introduction}

Unsteady aerodynamics is increasingly prevalent in aeronautics and fluids dynamics research in the 21st century. Unsteady flow phenomena are present in a wide range of problems and in diverse fields. The study of biological flight based on flapping wings involving high lift and large-scale vortex shedding, for example, is of interest to both biologists and
engineers~\citep{eldredge2019leading}. Unsteady flows exhibit rapid changes in bound circulation of the lifting surface and apparent-mass effects, as well as nonlinear phenomena such as flow separation and vortex shedding~\citep[chap. 8]{leishman_heli} .

Theoretical formulations and low-order models for unsteady flows are typically based on the Boundary-Element-Method (BEM) approach. These solve the unsteady potential flow equations with farfield boundary conditions being naturally enforced, and require calculations only on
the modelled surface(s) and wake(s). Since unsteady flows involve
time-varying circulation and shedding of vorticity from surfaces, inviscid
vortex elements/sheets are used to model these
phenomena~\citep{darakananda2019versatile}. The ease of setup/use and
rapid solution times offered by these methods has made them
useful for initial design and analysis. 

Thin-aerofoil theory is derived from a simplification of the boundary-element-method where the aerofoil is assumed to be thin (consisting only of a camber-line), and the boundary condition is transferred from the camber line to the chord line~\citep{ramesh2020leading}. The vorticity distribution on the chord line is modelled by a general Fourier series, with a special ``$A_0$'' term that is infinite at the leading edge. This term represents the ``suction peak'' caused by the flow having to turn around the aerofoil leading edge when the stagnation point moves away from the leading edge. A Kutta condition is applied at the trailing edge to uniquely specify the net circulation about the aerofoil. Thin-aerofoil theory has an advantage over panel methods in providing closed-form expressions for forces and moment on the aerofoil, and physical interpretations for the Fourier coefficients. It may also be more accurate than panel methods for thin sections as it has no errors associated with geometric discretisation.

Originally developed for steady flows about an aerofoil at a constant
angle of attack, thin-aerofoil theory has been extended to
unsteady flows by~\citet{katz_plotkin_book_1991}.~\citet{kiran_journal1}
have derived such a theory valid for arbitrarily large amplitudes and
non-planar wakes. Though the need to model free vorticity interaction
in the wake makes this method semi-numerical in comparison with
completely closed-form theory like~\citet{theodorsen1935}, it is
applicable to a wider range of scenarios occurring in nature and
engineering. This method has been used in studies on diverse topics
including design of efficient flapping aerofoils through
morphing~\citep{willis2014multiple}, and power generation by
self-sustained oscillation of an aeroelastic
aerofoil~\citep{ramesh2015limit}. The ``$A_0$'' term in this method is
of particular interest and has been used to develop a new aerodynamic
entity called the Leading-Edge Suction Parameter (LESP)
in~\citet{ramesh2014discrete}. It has been shown through experimental
and numerical verification that the process of leading-edge vortex
(LEV) formation is strongly correlated to the local suction at the
leading edge and hence the
LESP~\citep{ramesh2018leading,deparday2018critical,deparday2019modeling}. This
parameter has been used to predict and control the occurrence of LEV
formation, and in discrete-vortex methods for modelling intermittent
LEV shedding on aerofoils/wings where the LESP is used to both predict
LEV formation and modulate the strength of discrete vortices shed from
the leading
edge~\citep{ramesh2014discrete,ramesh2017experimental,hirato2019vortex}.

Despite these successful applications listed above, the typical implementation of the Kutta condition in unsteady thin-aerofoil theory (UTAT) suffers from a limitation, as shown later in the article.~\citet{roesler2018discretization} have derived discretisation requirements for the unsteady vortex-lattice method (also known as the lumped-vortex method for 2D problems) in terms of time-step, bound vortex spacing, and the gap between the trailing-edge and last-shed discrete wake vortex. Using the Wagner analytical solution as a reference, they showed that incorrect discretisation led to an incorrect implementation of the Kutta condition, resulting in errors in force prediction. This article provides a similar contribution for unsteady thin-aerofoil theory. 

Inviscid boundary-element methods require, in addition to the surface tangency condition, a Kutta condition to be applied at the trailing edge to uniquely specify the net circulation about the aerofoil. This condition provides an additional boundary condition that represents the consequence of viscosity at the trailing edge in a real fluid flow. In essence, the Kutta condition allows selection of the correct solution among the many possible solutions. There are many ways in which the Kutta condition has been specified in the aerodynamics literature. The common thread between them is that they all require the flow to leave smoothly from the trailing edge of the aerofoil~\citep{xia2017unsteady}. 

For steady flows,~\citet{poling1986response} have expressed the Kutta condition in a number of forms: (i) pressure difference (between aerofoil upper and lower surfaces) at the trailing edge is zero, (ii) velocity difference at the trailing edge is zero, (iii) no vorticity shedding from trailing edge. For unsteady flows,~\citet{basu1978unsteady} argue that only form (i), wherein the pressure difference is zero at the trailing edge, is valid.~\citet{jones2003separated} derived an analytical Kutta condition for unsteady flow past a flat plate which stated that, (i) vortex sheet is continuous across the trailing edge (between bound and wake vorticity), (ii)  average velocity is continuous across the trailing edge. 

From the references cited above, we see that finite velocity at the trailing edge (or flow leaving smoothly from the trailing edge) is a necessary but insufficient criterion for unsteady flows. In these scenarios, where vorticity is shed continuously from the trailing edge, an additional condition is required to specify the Kutta condition. This may be expressed as vorticity being non-zero and continuous across the trailing edge into the wake, or as pressure difference at the trailing edge being equal to zero.

In the unsteady vortex lattice method (UVLM)~\citep{roesler2018discretization,katz_plotkin_book_1991}, the bound vorticity on the aerofoil and shed vorticity in the wake are both modelled by discrete vortex singularities. The boundary condition is enforced at a set of control points, and the solutions for the unknown bound vortex strengths and last-shed wake vortex are obtained by solving a linear system of equations. The first part of the Kutta condition, requiring a finite velocity at the trailing edge, is trivially satisfied by not placing a discrete bound vortex at the trailing edge (since the location of a vortex singularity has infinite velocity). The boundary condition is not enforced at the trailing edge point, and the solution at this location is obtained through interpolation. The second part of the Kutta condition is not explicitly specified, and it is possible to obtain an incorrect solution based on the choice of discretisation. The discretisation requirements specified by~\citet{roesler2018discretization} are hence an indirect way of applying the second part of the Kutta condition which ensures that the pressure difference at the trailing edge is zero. 

In unsteady thin-aerofoil theory (UTAT)~\citep{katz_plotkin_book_1991,kiran_journal1}, the bound vorticity is represented as a continuous vortex sheet using a Fourier series. The boundary condition is integrated over the aerofoil chord, resulting in a linear equation with only one unknown, the strength of the last-shed wake vortex. The Kutta condition is typically said to be enforced by the form of the Fourier series, which is such that its value evaluated at the trailing edge is zero. There are however two concerns to note regarding this statement. Though the point-wise value of the infinite Fourier series at the trailing edge is zero, the value as the edge is approached isn't necessarily zero~\citep{rienstrakutta}. Additionally, as we have observed earlier in the article, vorticity being zero at the trailing edge (and hence velocities on upper and lower surface at the trailing edge being equal) is a condition for steady flow and isn't true for unsteady flow. In the typical UTAT formulation, the shed wake vorticity is represented by a point vortex or regularised vortex blob. There is hence a discontinuity in the vortex sheet across the trailing edge, resulting in the second part of the Kutta condition not being satisfied.

In the new approach proposed in this article, the shed wake at any discrete time step in unsteady thin-aerofoil theory is represented by a continuous distribution of vorticity derived from the exact Wagner solution rather than by a point vortex or regularised vortex blob. We show that this results in vorticity being continuous across the trailing edge, and consequently, velocities being finite and pressure difference going to zero at the trailing edge.

\section{Theory and Methods}

\subsection{Unsteady Thin-Aerofoil Theory - general formulation}

A general summary of unsteady thin-aerofoil theory is given below before the original and modified methods of implementing the Kutta condition are discussed. A more detailed description may be found in~\cite{katz_plotkin_book_1991}  and \cite{ramesh2020leading}. 

\begin{figure}
 \centering{
  \includegraphics[width=0.98\textwidth]{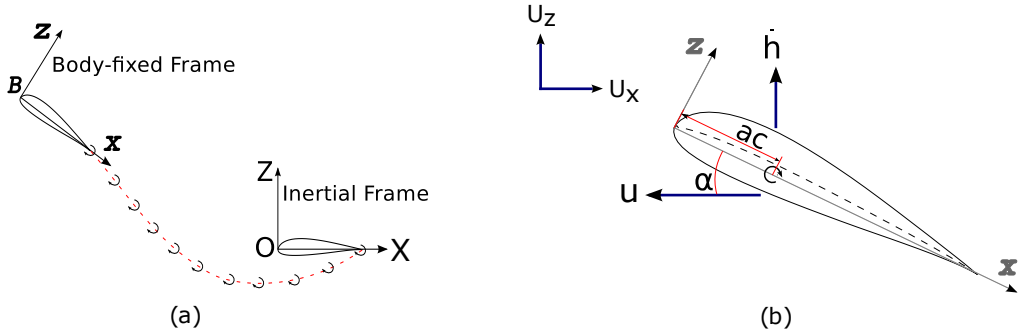}
  \caption{(a) Illustration of unsteady thin-aerofoil theory and the
    time-stepping method, (b) aerofoil and freestream velocities
    (positive as shown) and pitch-axis location.}
\label{fig:timestepping}}
\end{figure}

The method is valid for
arbitrary variations in freestream velocity and aerofoil kinematics,
and contains no assumptions of small motion amplitudes or planar wakes
(which are necessary in fully closed-form
theories). Figure~\ref{fig:timestepping}(a) illustrates the method, with
the inertial reference frame given by $OXYZ$ and the body frame
(attached to the moving aerofoil) by $Bxyz$. The two frames coincide
at time $t=0$ and at each time step, wake vorticity is shed from the trailing-edge.

 The vorticity distribution over the aerofoil at any time step, $\gamma(x,t)$, is modelled as a Fourier series,

\begin{equation}
\gamma(\theta,t)=2U_{ref}\left[A_{0}(t)\frac{1+\cos \theta}{\sin \theta}
  + \sum_{n=1}^{\infty}A_{n}(t) \sin(n\theta)\right]
\label{eqn:tat_vort}
\end{equation}

\noindent where $A_{0}(t)$, $A_{1}(t)$, ..., $A_{n}(t)$ are time-dependent Fourier coefficients, $c$ is the aerofoil chord, $U_{ref}$ is a reference velocity for nondimensionalisation chosen according to the problem, and $\theta$ is a variable of transformation related to the chordwise coordinate as 

\begin{equation}
x=\frac{c}{2}(1-\cos\theta)
\end{equation}

The Fourier coefficients are determined from the instantaneous normal downwash, $W(x,t)$, by integrating the zero-normal-flow boundary condition on the chord line. 

\begin{equation}
A_{0}(t) = -\frac{1}{\pi}
\displaystyle \int_{0}^{\pi}\frac{W(x,t)}{U_{ref}}d \theta , \qquad
A_{n}(t) =
\frac{2}{\pi} \displaystyle \int_{0}^{\pi}\frac{W(x,t)}{U_{ref}} \cos n
\theta d \theta .
\label{eqn:fourier_terms}
\end{equation}

Using the general formulation of~\citet{kiran_journal1} which removes small-angle approximations in the theory, the normal downwash on the aerofoil camber-line transferred to the chord-line is determined from the aerofoil, freestream and induced velocities (shown in figure~\ref{fig:timestepping}(b)) as 

\begin{multline}
W(x,t) = \frac{\partial \eta_c}{\partial x}\left((U_X + u)\cos \alpha +
(\dot{h} - U_Z) \sin \alpha + \frac{\partial \phi_{w}}{\partial x}\right) \\
- (U_X + u) \sin \alpha - \dot{\alpha}(x-ac)
  + (\dot{h} - U_Z) \cos \alpha - \frac{\partial \phi_{w}}{\partial z}
\label{eqn:fourier2}
\end{multline}

\noindent where $\eta_c(x)$ is the camber variation on the aerofoil,
$a$ is the nondimensional location of the pitch axis from $0-1$, and
$\partial \phi_{w}/\partial x$ and $\partial \phi_{w}/\partial z$ are velocities induced on the camber-line in directions tangential and normal to chord by shed vorticity in the wake. Arbitrary motion kinematics of the aerofoil in the 2D plane are
represented by the time-varying parameters: plunge velocity in the
$Z$ direction $\dot{h}(t)$, horizontal velocity in the negative $X$ direction $u(t)$, and pitch angle $\alpha(t)$. The velocities $U_X(t)$ and
$U_Z(t)$ are horizontal and vertical components of external freestream flow which may also result from gusts or other perturbations.

The Vatistas regularized vortex model~\citep{vatistas1991simpler} is used to calculate the velocity induced by discrete vortex blobs on the camber-line. The velocities induced at a location $(X,Z)$ by a vortex blob at $(X_w,Z_w)$ with circulation $\Gamma_w$ and vortex core radius $v_{core}$ are given by

\begin{eqnarray}
u &=& \frac{\Gamma_w}{2 \pi} \frac{Z-Z_w}{\sqrt{((X-X_w)^2+(Z-Z_w)^2)^2+v_{core}^4}} \nonumber\\
w &=& -\frac{\Gamma_w}{2 \pi} \frac{X-X_w}{\sqrt{((X-X_w)^2+(Z-Z_w)^2)^2+v_{core}^4}}
\label{eqn:induced}
\end{eqnarray}

Kelvin's circulation theorem gives, 

\begin{equation}
 \Gamma_B(t) - \Gamma_B(t- \Delta t) + \Gamma_w = 0
  \label{eqn:kelvin}
\end{equation}

\noindent where $\Gamma_w$ is the circulation of the last-shed wake vortex, and $\Gamma_B$ is the bound circulation calculated by
integrating the vorticity distribution (eqn.~\ref{eqn:tat_vort}) over
the aerofoil chord.

\begin{equation}
  \Gamma_B(t) = U_{ref}c\pi \left(A_0(t) + \frac{A_1(t)}{2} \right)
    \label{eqn:Gamma}
\end{equation}

For further discussion in this article, with the aim of simplifying the maths, we assume without any loss of generality that the aerofoil has no camber, that it  travels at a constant horizontal velocity and that there are no external disturbances. The reference velocity is taken as the
horizontal velocity.

\begin{equation}
  u(t) = U_{ref} = u, \qquad U_X=U_Z=0, \qquad \eta_c(x) = 0
\end{equation}

The Fourier coefficients may be written as the sum of 2 components $A_{n} = A_{n_m}+A_{n_i}$ where the subscript $i$ denotes the contribution of the last-shed wake vortex (unknown at each time step), and the subscript $m$ denotes contributions from the aerofoil kinematics, freestream velocity and all wake vorticity shed in previous time steps (known at each time step).  

From eqns.~\ref{eqn:fourier_terms} and~\ref{eqn:fourier2},

\begin{align}
    A_{0_m} &= \sin \alpha - \frac{\dot{\alpha}c}{u}\left(a-\frac{1}{2}\right)
  - \frac{\dot{h}}{u} \cos \alpha + \frac{1}{u\pi} \int_0^\pi \left(\frac{\partial \phi_{w}}{\partial z}\right)_{p} d\theta  \nonumber \\
    A_{1_m} &= \frac{\dot{\alpha}c}{2u} - \frac{2}{u\pi} \int_0^\pi \left(\frac{\partial \phi_{w}}{\partial z}\right)_{p} \cos \theta d\theta  \nonumber \\
    A_{2....n_m} &= - \frac{2}{u\pi} \int_0^\pi \left(\frac{\partial \phi_{w}}{\partial z}\right)_{p} \cos n \theta d\theta  
    \label{eqn:fourier_motion}
\end{align}

\noindent where the subscript $p$ in the integrals denotes that this term is calculated from vortices shed in previous time steps based on equation \ref{eqn:induced}. Using eqns.~\ref{eqn:kelvin} and \ref{eqn:Gamma}, the solution at any time step is obtained from, 

\begin{equation}
 U_{ref}c\pi \left(A_{0_m}(t) + A_{0_i}(t) + \frac{A_{1_m}(t)}{2} + \frac{A_{1_i}(t)}{2} \right) + \Gamma_w - \Gamma_B(t- \Delta t) = 0
 \label{eqn:solve}
\end{equation}

\begin{figure}
 \centering{
  \includegraphics[width=0.48\textwidth]{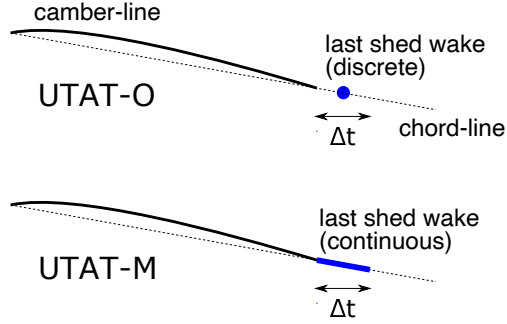}
  \caption{Illustration of the difference in modeling of the last-shed wake between UTAT-O (original) and UTAT-M (modified)}
\label{fig:utats}}
\end{figure}

As $A_{0_i}$ and $A_{1_i}$ are calculated from $\Gamma_w$ and all the other terms are known, $\Gamma_w$ (last-shed wake vorticity) is the only unknown at any given time step. The determination of $A_{0_i}$ and $A_{1_i}$ in the original and modified algorithms of UTAT are discussed in sections~\ref{sec:utat_orig} and~\ref{sec:utat_mod} respectively. These methods are labelled as "UTAT-M" (modified) and "UTAT-O" (original). Figure~\ref{fig:utats} shows the difference in modelling of the last-shed wake vorticity between them. Once the solution is obtained at any given time step, the pressure difference coefficient on the aerofoil may be evaluated from the Fourier coefficients as~\citep{ramesh2020leading}  

\begin{multline}
  \Delta C_p = 4\left(\cos \alpha + \frac{\dot{h}}{u} \sin \alpha + \frac{1}{u} \frac{\partial \phi_{w}}{\partial x}\right) \\ 
  \left[A_0\left(\frac{2 \sin (\theta/2)}{r+2\sin^2(\theta/2)} - \tan(\theta/4)\right) + \sum_{1,2,...n}^{n} A_n \sin n \theta \right] \\
  + 2\frac{c}{u}\left[\dot{A_0}(\theta +\sin \theta) + \dot{A_1}\left(\frac{\theta}{2}-\frac{\sin 2 \theta}{4}\right)   + \sum_{2,3,...n}\frac{\dot{A_n}}{2}\left(\frac{\sin(n-1)\theta}{n-1} - \frac{\sin (n+1)\theta}{n+1} \right) \right] 
\label{eqn:p_comb}
\end{multline}

\noindent where $r$ is the leading-edge radius nondimensionalised with respect to aerofoil chord.

\subsection{Original algorithm}    
    \label{sec:utat_orig}
The last-shed wake vorticity in UTAT is typically modelled with a singular discrete vortex or regularised vortex blob~\citep{katz_plotkin_book_1991,kiran_journal1}. It is placed aft of the chord at a distance half that travelled by the aerofoil in a given time step ($u\Delta t/2$). The Fourier coefficients resulting from this vortex are calculated as 

\begin{align*}
    A_{0_i} &= \frac{1}{u\pi} \int_0^\pi \left(\frac{\partial \phi_{w}}{\partial z}\right)_{c} d\theta  \\
    A_{1....n_i} &= - \frac{2}{u\pi} \int_0^\pi \left(\frac{\partial \phi_{w}}{\partial z}\right)_{c} \cos n \theta d\theta \numberthis \label{eqn:original_fourier}
\end{align*}

\noindent where the subscript $c$ in the integrals denotes that this term is calculated from the vortex shed in the current time step based on equation \ref{eqn:induced}. Equation~\ref{eqn:solve} is now solved for $\Gamma_w$, as a linear equation  or using Newton iteration. 

\subsection{Modified algorithm}
 \label{sec:utat_mod}
 
Here, we model the last-shed wake vorticity at any time step using a continuous distribution. Seeing as each time step is essentially a step change in the kinematic conditions, this continuous distribution and the aerofoil downwash due to it are derived from Wagner's exact solution.~\citet{epps2018vortex} have derived the vortex sheet strength for Wagner's problem, resulting from noncirculatory, circulatory and wake effects. Expressed in our chordwise transformation variable $\theta$, the bound vorticity distribution due to downwash induced by the wake vorticity is,

\begin{equation}
    \gamma_i(\theta,t) = -2 W_0 \left[R_{0}(t^*)\frac{1 + \cos \theta}{\sin \theta}
  - 2 \sum_{n=1}^{\infty}(-1)^n R_{n}(t^*) \sin(n\theta)\right]
  \label{eqn:induced_vort}
\end{equation}

\noindent where $t^*=U_{ref}t/c$ is nondimensional time, $W_0$ is the magnitude of the step change in downwash at the $3/4$-chord location, and 

\begin{align}
    R_n(t^*) &= \frac{1}{2\pi}\int_{-\infty}^\infty Q_n(k) S(k) e^{ik t^*} dk \\
    Q_n(k) &= \int_0^\infty e^{-ik \cosh k} e^{-n\ \zeta} d \zeta \\
    S(k) &= \frac{1/ik}{K_1(ik) + K_0(ik)}
\end{align}

$R_0(t^*)$ and $R_1(t^*)$ have closed-form expression given by

\begin{align}
R_0(t^*) &= 1 - \Phi(t^*) \\ 
R_1(t^*) &= \Phi(t^*) - \Psi(t^*)
\end{align}

\noindent where $\Phi(t^*)$ and $\Psi(t^*)$ are the Wagner's and Kussner's functions, respectively. Comparing eqn.~\ref{eqn:induced_vort} with eqn.~\ref{eqn:tat_vort} yields the value of the Fourier coefficients resulting from wake vorticity in Wagner's solution. The Fourier coefficients representing the induced bound vorticity for a step change in kinematics occurring in a time step $\Delta t$ are hence 

\begin{align}
    A_{0_i} &= -\frac{W_0}{u}\left(1 - \Phi(\Delta t^*)\right) \nonumber \\
    A_{1_i} &= -2\frac{W_0}{u}\left(\Phi(\Delta t^*) - \Psi(\Delta t^*)\right) \nonumber \\
    A_{n_i} &= (-1)^n 2\frac{W_0}{u} R_n(\Delta t^*)
    \label{eqn:wagner_fourier}
\end{align}

The wake vorticity in Wagner's theory resulting from the step change $W_0$ is derived from ~\citet{epps2018vortex} as 

\begin{align}
\gamma_w &= -\pi c W_0 \Psi ^\prime(t^*), \qquad 0 \le t^* \le \Delta t^* \label{eqn:wagner_wake_dist} \\ 
\Gamma_w &= -\pi c W_0 \Psi (\Delta t^*)
\label{eqn:wagner_wake}
\end{align}

Using eqns.~\ref{eqn:wagner_fourier} and \ref{eqn:wagner_wake} in eqn.~\ref{eqn:solve}, the problem is reduced to solving for $W_0$ which is the unknown change in 3/4-chord downwash at any time step. We obtain the simple linear equation, 

\begin{equation}
    W_0(t) = U_{ref}\left(A_{0_{m}}(t) + \frac{A{1_{m}}(t)}{2}\right) - \frac{\Gamma_B(t- \Delta t)}{c\pi}
\end{equation}

The complete solution is then calculated from eqns.~\ref{eqn:wagner_fourier},~\ref{eqn:fourier_motion} and~\ref{eqn:solve}. After obtaining the solution at each time step, the continuous wake vorticity is converted into a vortex blob located at the centroid of the distribution with circulation obtained from eqn.~\ref{eqn:wagner_wake}.  

\iffalse

\begin{equation}
    U_{ref} c \pi (A_{0_{m}} + A{_{1_{m}}}/2 - W_0(1 - \Psi(t^*))) - \pi c   W_0 \Psi(t^*) + \sum \Gamma_{TEV,prev} = 0
\end{equation}

\noindent the problem is reduced to solving for $W_0$ which is the unknown "effective change in 3/4-chord downwash" at any time step. We obtain the simple linear equation, 

\fi

\section{Results and verification}

Results from the modified implementation of UTAT detailed above are compared against those from the original implementation in this section. These methods are labelled as "UTAT-M" (modified) and "UTAT-O" (original), respectively. CFD solutions of the incompressible Navier-Stokes equations, and Theodorsen's and Wagner's exact analytical solutions are used as "true solutions" for verification.

The Euler CFD simulations are carried out using the open-source CFD toolbox OpenFOAM. This setup has previously been used in~\citet{ramesh2020leading} and~\citet{bird2021unsteady} to verify 2D and 3D analytical solutions in unsteady aerodynamics. A body-fitted computational mesh is moved in accordance with prescribed rate laws, and the time-dependent governing equations are solved using a finite volume method.  A second-order backward implicit scheme is adopted to discretise the transient terms, and second-order Gaussian integration schemes with linear interpolation for the face-centred values of the variables are used for the gradient, divergence and Laplacian terms. Viscosity is set to zero, and a slip boundary condition is used for the moving aerodynamic surface. A NACA0004 section is considered, in order to best match the thin-aerofoil assumption in theory. An O-mesh is constructed with 360 cells around the aerofoil, 252 cells in the wall-normal direction, and with the farfield extending to 25 chord lengths in all directions from the aerofoil.

\subsection{Recovery of Wagner's solution}

\begin{figure}
\centering
    \includegraphics[width=0.5\textwidth]{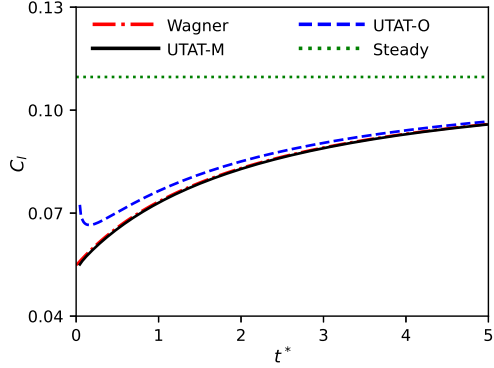}

	\caption{Comparison of lift coefficient from the two formulations of UTAT against Wagner's exact solution for the case of a $1$ deg step change in pitch angle. The steady state solution ($2\pi \alpha$) is also shown.}
	\label{fig:wagner}	
\end{figure}

Figure~\ref{fig:wagner} shows lift coefficients from UTAT-M and UTAT-O compared against Wagner's solution for a step change of $1$ deg in pitch angle. UTAT-M is seen to converge with the exact solution from the start of the simulation, whereas UTAT-O has an error that decreases with time. This shows that the error associated with an incorrect Kutta condition increases with increasing reduced frequency.  

\begin{figure}
\centering
    \includegraphics[width=0.8\textwidth]{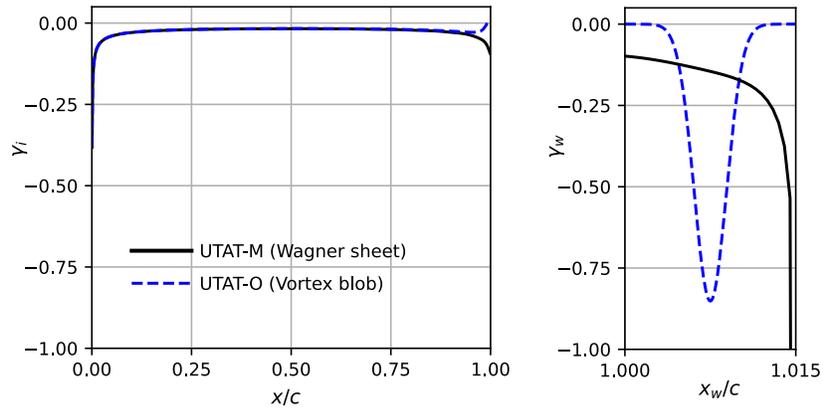}
	\caption{Induced bound vorticity ($\gamma_i$) and shed wake vorticity ($\gamma_w$) for a $1$ deg step change in pitch angle, occurring in $\Delta t^* = 0.015$. }
	\label{fig:methods_compare}	
\end{figure}

To understand the reason behind the error in UTAT-O, we compare the solutions from the two formulations after the very first time step. Figure~\ref{fig:methods_compare} compares the last-shed wake vorticity distribution ($\gamma_w$, calculated from equation~\ref{eqn:wagner_wake_dist}) and the resulting induced bound vorticity distribution ($\gamma_i$, calculated from equations~\ref{eqn:tat_vort} and~\ref{eqn:wagner_fourier}) from the two methods after the first time step ($\Delta t^* = 0.015$). In the original formulation (UTAT-O), we see that $\gamma_i$ goes to zero at the trailing edge. The last-shed wake vorticity, modelled as a vortex blob, is placed behind the trailing edge at a distance equal to half that travelled by the aerofoil during the time step. In the modified formulation, the distribution of shed vorticity in the wake (eqn.~\ref{eqn:wagner_wake_dist}) and the induced bound vorticity (eqn.~\ref{eqn:wagner_fourier}) are both derived from Wagner's exact solution. Here, we see that that $\gamma_i$ and $\gamma_w$ are non-zero at the trailing edge, as expected in the case of unsteady flow. They are also continuous across the trailing edge, resulting in no discontinuity/singularity at the trailing edge. Therefore both parts of the unsteady Kutta condition are satisfied in UTAT-M; the velocities are finite at the trailing edge, and the bound vortex sheet is non-zero and continuous across the trailing edge and into the wake. UTAT-O only satisfies the first part of the Kutta condition, resulting in an erroneous solution.

\iffalse
Differences between the two methods are illustrated in Fig.~\ref{fig:methods_compare}, considering a step change in pitch angle equal to $1$ deg in a time step $dt^*=0.015$. Part (a) of the figure shows the difference in modelling of the last-shed wake vorticity between the two methods. Part (b) shows the bound vorticity induced by the last-shed wake vorticity on the aerofoil chord ($\gamma_i$), and the distribution of the last-shed vorticity in the wake ($\gamma_w$). 
\fi

\begin{figure}
\centering
    \includegraphics[width=0.5\textwidth]{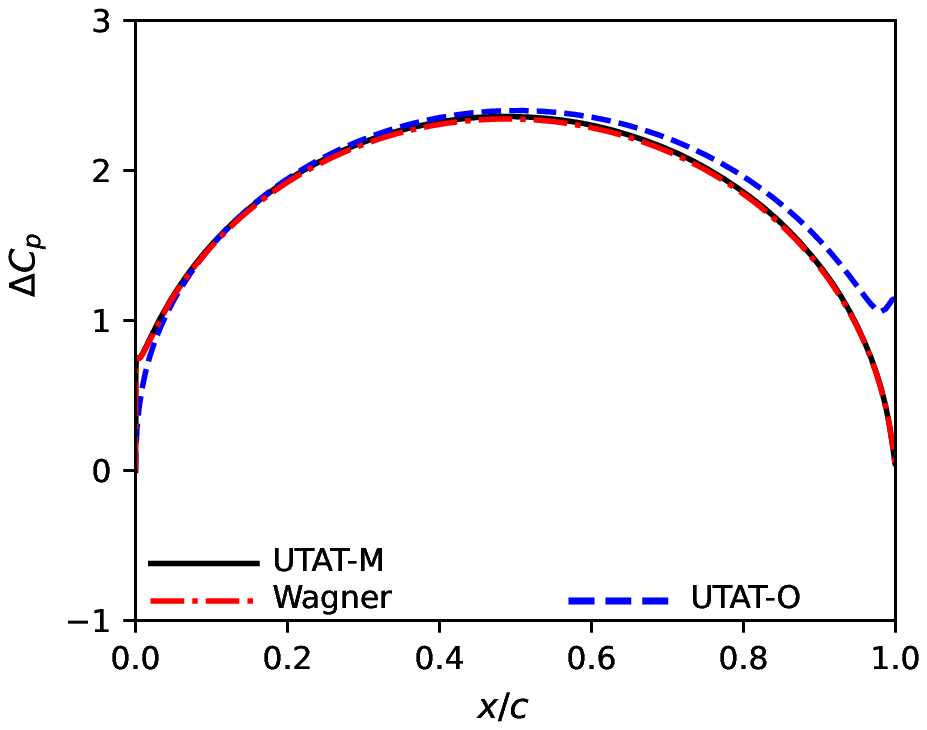}
    \Put(-153,130){\includegraphics[width=0.31\textwidth]{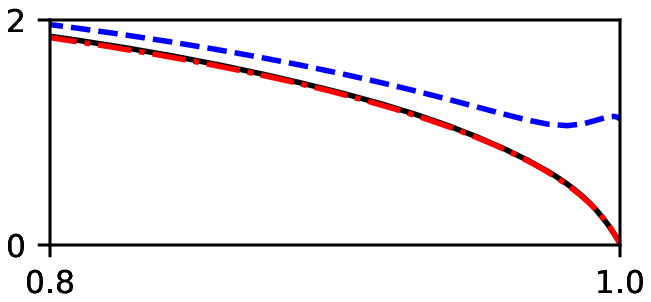}}
	\caption{Comparison of pressure difference coefficient from the two formulations of UTAT against Wagner's exact solution for the case of a case of step change in pitch angle equal to $1$ deg, after the first time step at $t^*=0.015$. Inset shows zoomed-in pressure difference at trailing edge from 80\% chord.}
	\label{fig:wagner_delcp}	
\end{figure}

To further demonstrate that the second part of the Kutta condition is not satisfied in UTAT-O, we compare the pressure difference coefficients from the two methods against Wagner's exact solution after the first time step in figure~\ref{fig:wagner_delcp}. The inset in the plot shows the zoomed-in pressure difference coefficient near the trailing edge. The exact solution and UTAT-M have pressure difference going to zero at the trailing edge, while UTAT-O does not.

\subsection{Harmonic heave manoeuvre}

\begin{figure}
\centering
    \includegraphics[width=0.5\textwidth]{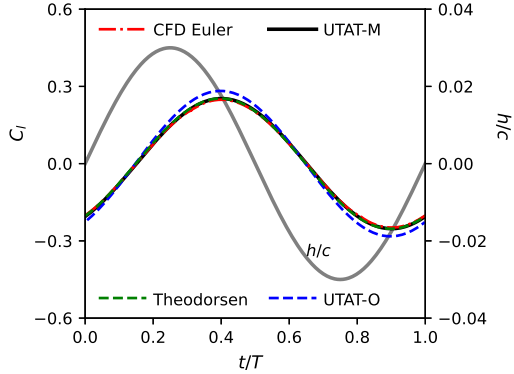}
	\caption{Comparison of lift coefficient from the two formulations of UTAT against Theodorsen's exact solution and an Euler CFD simulation for the case of a harmonic heave manoeuvre with $h_0/c = 0.03$, $k = 1.0$.}
	\label{fig:heave}	
\end{figure}

Next, a sine function of plunge is considered, $h = h_0 \sin \omega t$, with $h_0/c = 0.03$ and $k = 1.0$. Lift coefficients from the original and modified UTAT are compared against those from Theodorsen's theory and Euler CFD in Fig.~\ref{fig:heave}. We observe that the outlier is UTAT-O which has an error. 

\begin{figure}
  %\vspace{0.2in}
  \centering{
  \begin{minipage}{\textwidth}
    \centering{
      \begin{minipage}{0.49\textwidth}
        \centering{
          \includegraphics[width=0.99\textwidth]{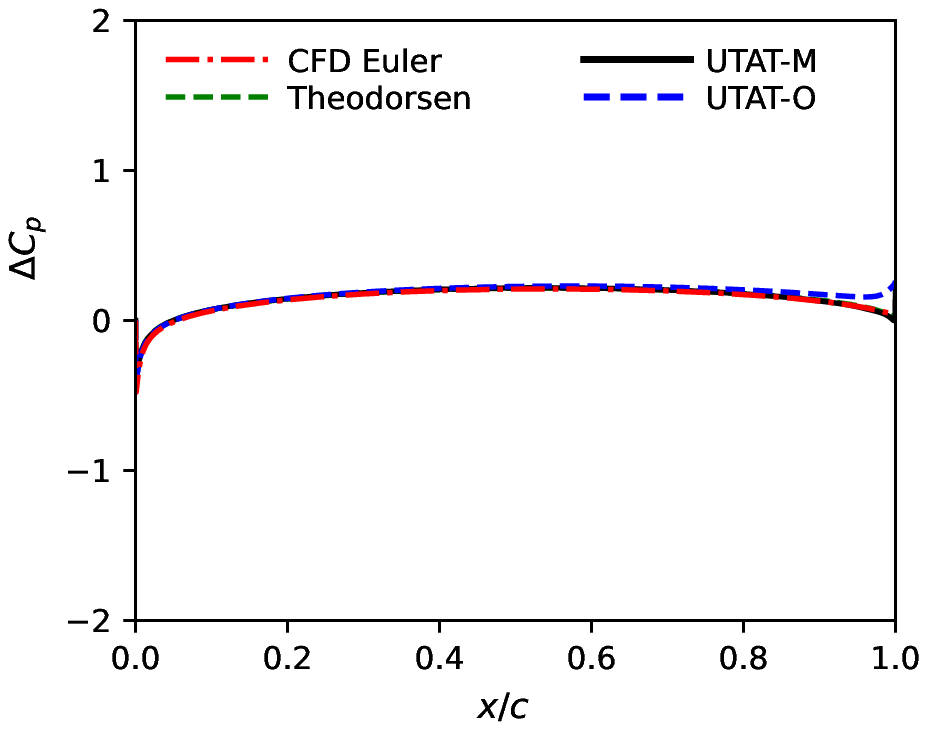}
          \Put(-50,120){\includegraphics[width=0.6\textwidth]{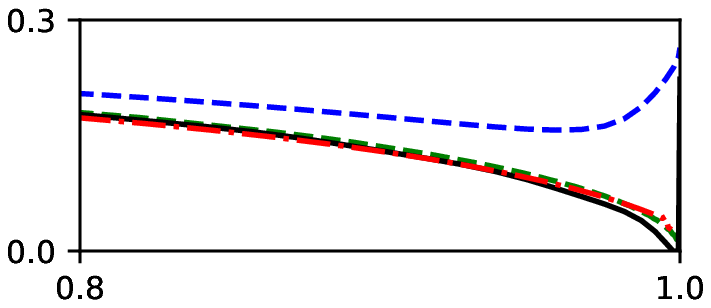}}
        }
      \end{minipage}
      \hfill
      \begin{minipage}{0.49\textwidth}
        \centering{
          \includegraphics[width=0.99\textwidth]{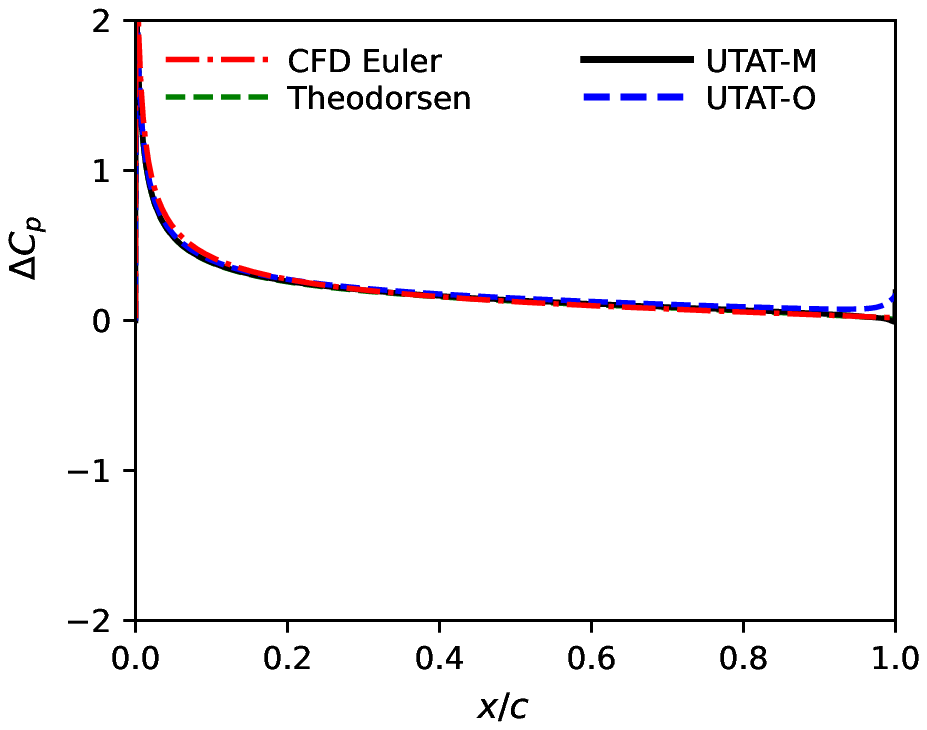}
          \Put(-50,120){\includegraphics[width=0.6\textwidth]{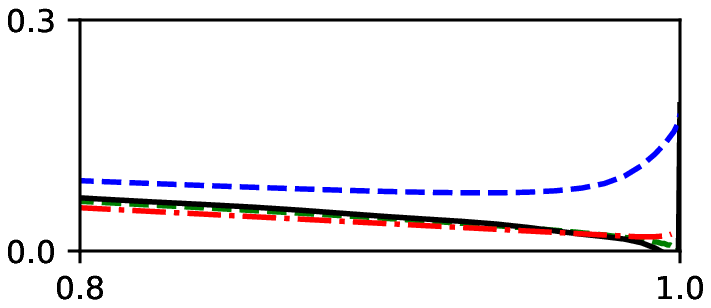}}
        }
    \end{minipage}}   
  \end{minipage}}
 
  \caption{Comparison of pressure difference coefficient from the two formulations of UTAT against Theodorsen's exact solution and an Euler CFD simulation for the case of a harmonic heave manoeuvre with $h_0/c = 0.03$, $k = 1.0$, at $t/T=0.25$ (left) and $t/T=0.5$ (right). Insets show zoomed-in pressure difference at trailing edge from 80\% chord.}
  \label{fig:delcp_h}
\end{figure}

To examine the reason for the error, we study the pressure-difference coefficient at equally spaced intervals during the motion $t/T = 0.25$ and $0.5$ in Fig.~\ref{fig:delcp_h}. The curves for $t/T = 0.75$ and $1.0$ are the exact negatives of these owing to the kinematic symmetry in this case. The insets in these plots show the zoomed-in pressure difference near the trailing edge. While UTAT-M matches the true solutions for the full extent of the aerofoil, UTAT-O shows an error near the trailing edge with pressure difference at the trailing edge being non zero. This shows that the Kutta condition is not truly satisfied in this method despite finite velocity at the trailing edge.

\subsection{Smoothed pitch ramp manoeuvre}

Finally, a case with non-harmonic pitch motion is considered. The pitch variation is defined by the Eldredge function which produces a ramp motion with smoothed corner~\citep{eldredge_computation_pitchramp, granlund2013unsteady}.

\begin{equation}
\alpha = \frac{K}{a_s}\left[\frac{\cosh(a_s(t^*-t_1^*))}
  {\cosh(a_s(t^*-t_2^*))}\right]+\frac{\alpha_{0}}{2}, \qquad
  a_s = \frac{\pi^2K}{2\alpha_{0}(1-\sigma)}, \qquad
  t_2^* = t_1^*+\frac{\alpha_{0}}{2K}
\end{equation}

\begin{figure}
\centering
    \includegraphics[width=0.55\textwidth]{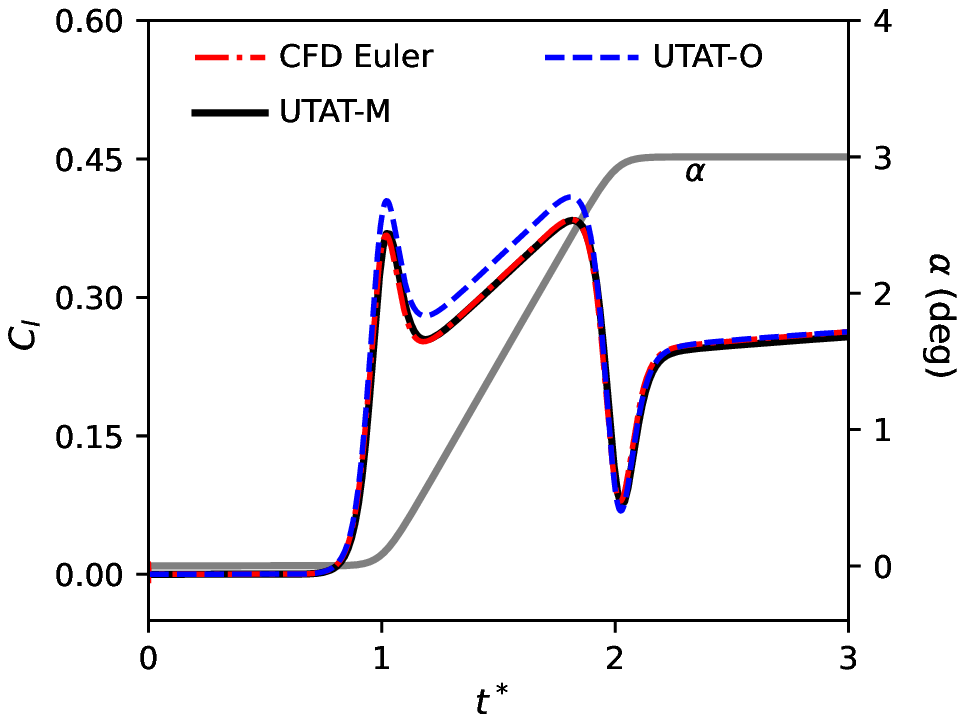}

	\caption{Comparison of lift coefficient from the two formulations of UTAT against an Euler CFD simulation for the case of a smoothed pitch ramp manoeuvre with $\alpha_0 = 3$ deg, $K = 0.026$, $\sigma=0.8$.}
	\label{fig:eld_cl}	
\end{figure}

\begin{figure}
  %\vspace{0.2in}
  \centering{
  \begin{minipage}{\textwidth}
    \centering{
      \begin{minipage}{0.49\textwidth}
        \centering{
          \includegraphics[width=0.99\textwidth]{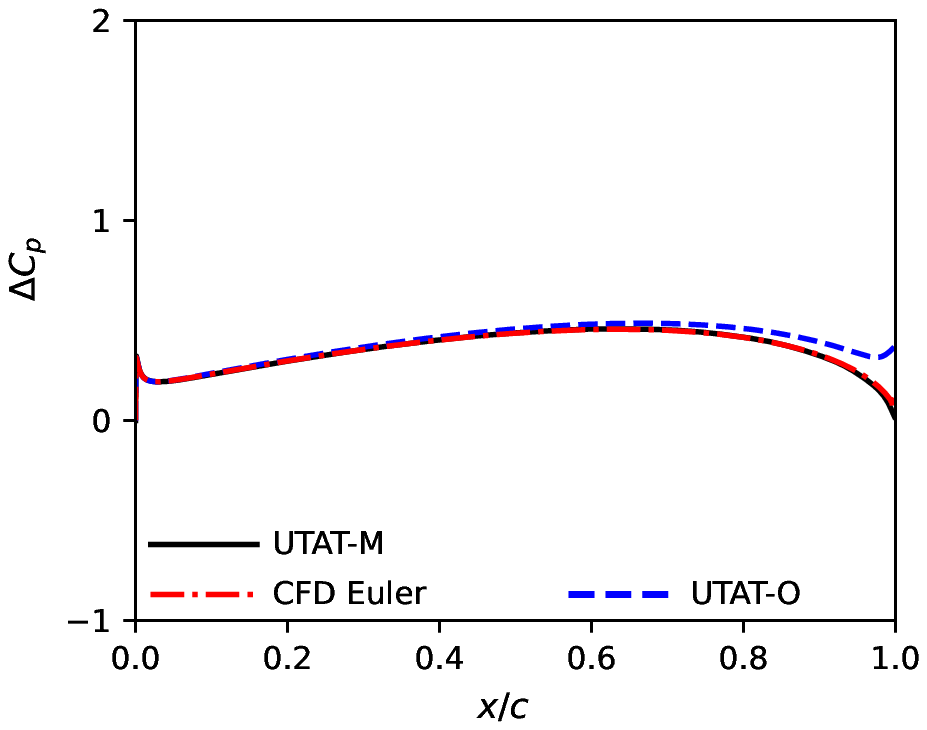}
          \Put(-50,230){\includegraphics[width=0.6\textwidth]{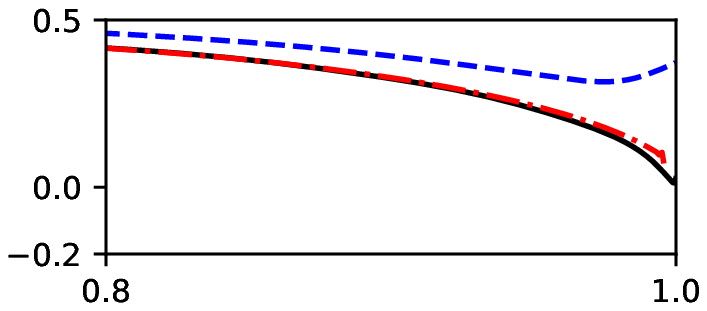}}
        }
      \end{minipage}
      \hfill
      \begin{minipage}{0.49\textwidth}
        \centering{
          \includegraphics[width=0.99\textwidth]{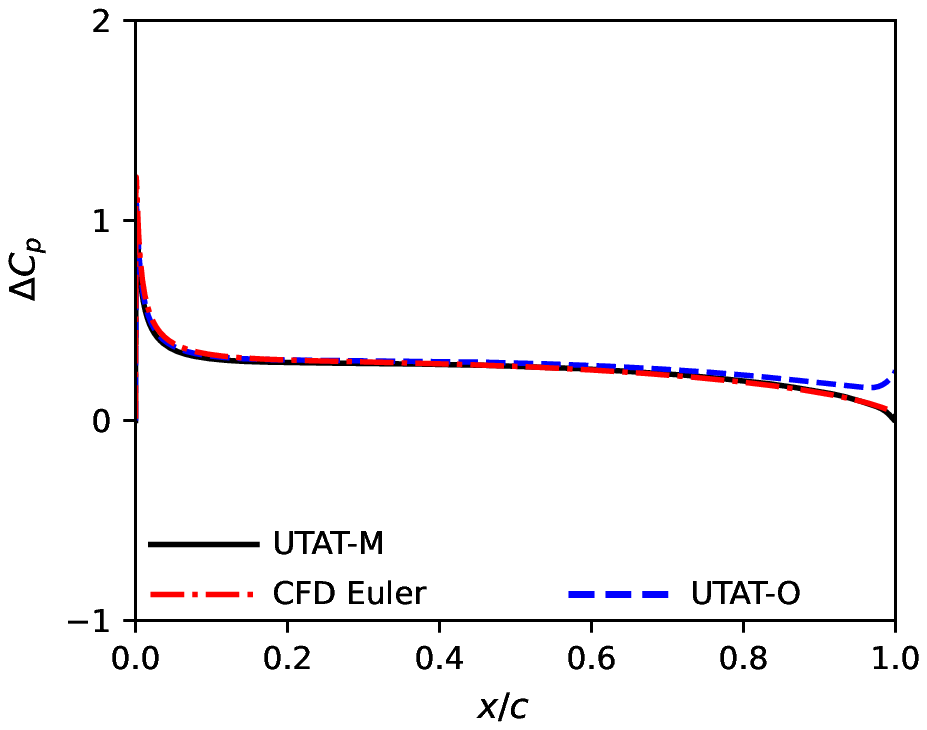}
          \Put(-50,230){\includegraphics[width=0.6\textwidth]{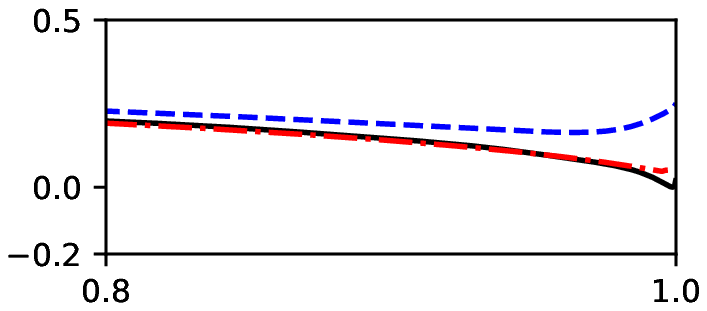}}
        }
    \end{minipage}}   
  \end{minipage}}
  \centering{
  \begin{minipage}{\textwidth}
    \centering{
      \begin{minipage}{0.49\textwidth}
        \centering{
          \includegraphics[width=0.99\textwidth]{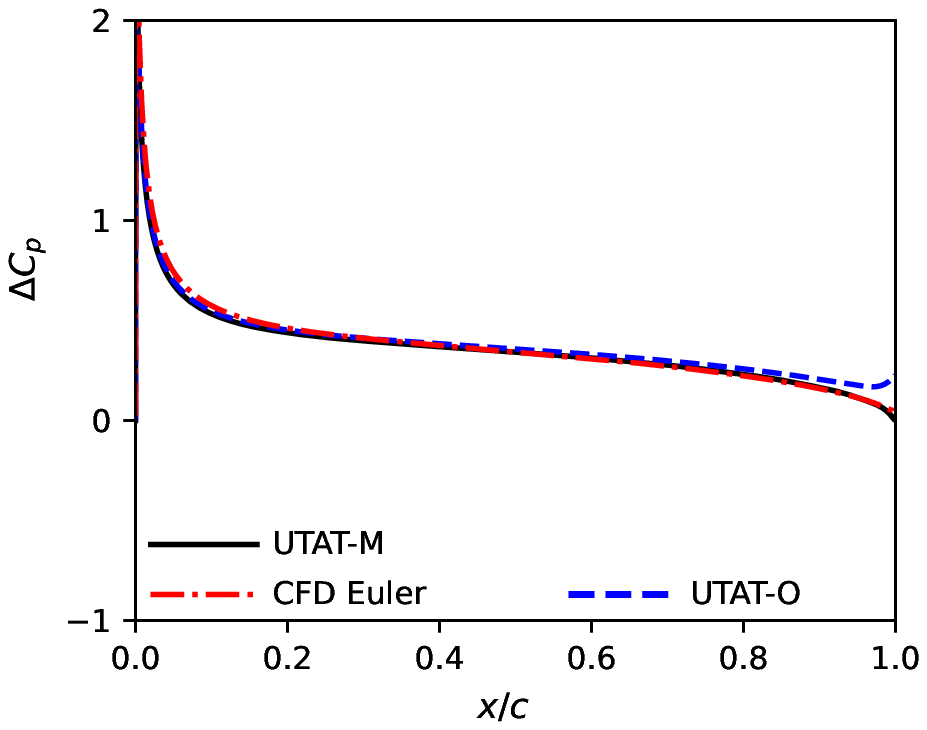}
          \Put(-50,230){\includegraphics[width=0.6\textwidth]{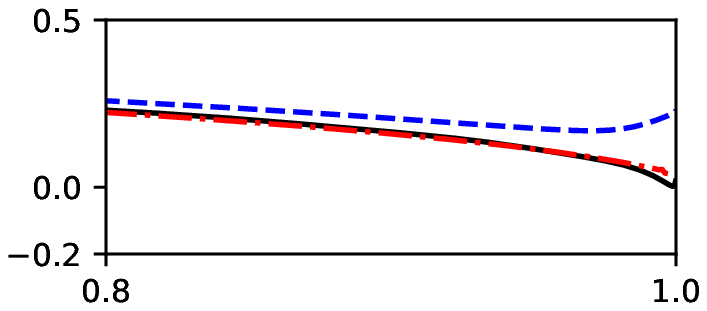}}
        }
      \end{minipage}
      \hfill
      \begin{minipage}{0.49\textwidth}
        \centering{
          \includegraphics[width=0.99\textwidth]{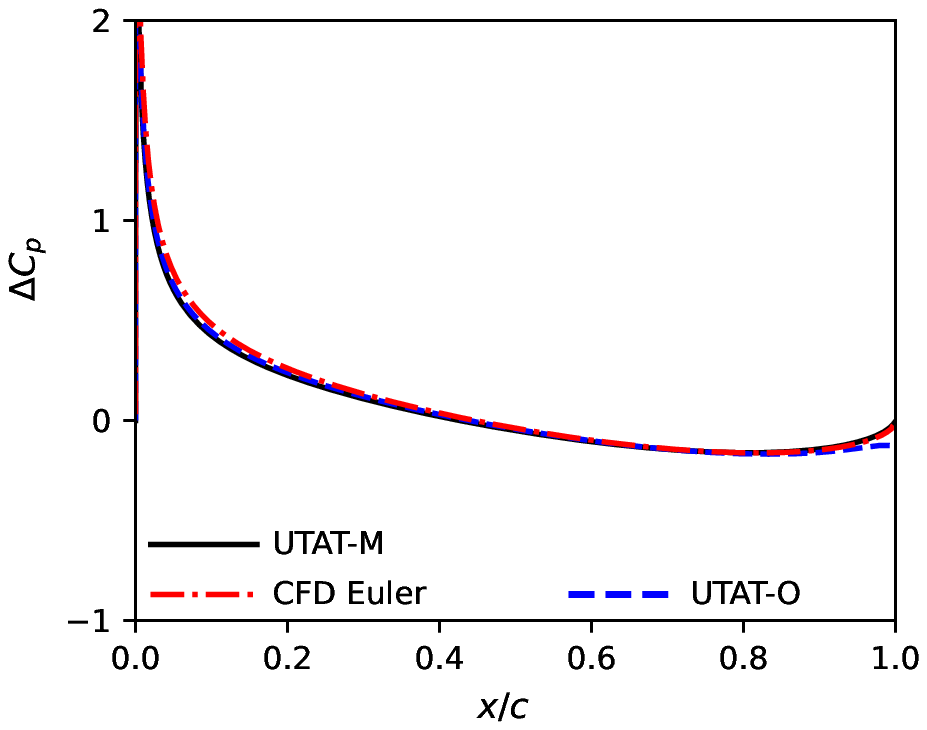}
          \Put(-50,230){\includegraphics[width=0.6\textwidth]{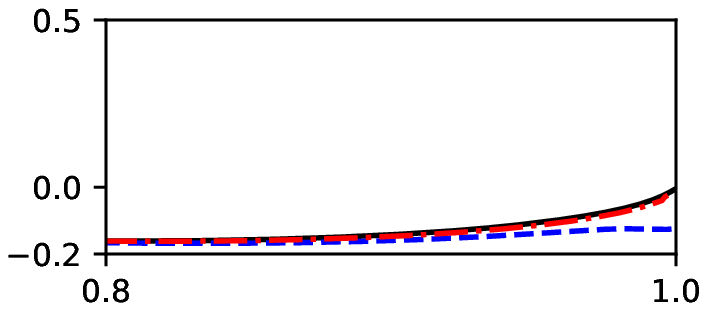}}
        }
    \end{minipage}}   
  \end{minipage}}
 
  \caption{Comparison of pressure difference coefficient from the two formulations of UTAT against an Euler CFD simulation for the case of a smoothed pitch ramp manoeuvre with $\alpha_0 = 3$ deg, $K = 0.026$, $\sigma=0.8$, at $t^*=1.0$ (top left), $t^*=1.25$ (top right), $t^*=1.75$ (bottom left)  and $t^*=2.0$ (bottom right). Insets show zoomed-in pressure difference at trailing edge from 80\% chord.
  }
  \label{fig:delcp_eld}
\end{figure}

Here, $t_1^*$ the nondimensional time at start of
ramp is taken as $1.0$. The parameter $\sigma$ is a nondimensional
measure of smoothing, set to $0.8$. $\alpha_{0}$ is the amplitude of
the ramp, equal to $3$ deg. $K$ is the reduced frequency of pitch
taken as $0.026$ so that the nondimensional ramp duration is
approximately equal to $1.0$. Pitch-axis is located at the leading
edge.

Lift coefficient comparison is shown in Fig.~\ref{fig:eld_cl}. While UTAT-M matches Euler CFD throughout the duration of the motion, UTAT-O shows an offset during the ramp-up phase. The error here is greater than in the previous case owing to the higher rate of motion in this case. Again, the pressure difference coefficients are used to illustrate the source of the error in figure~\ref{fig:delcp_eld}. We see that the original UTAT has an error near the trailing edge, and that the pressure coefficient doesn't go to zero at the trailing edge in this method. UTAT-M agrees with the Euler CFD solution, including near the trailing edge.

\section{Conclusions}

Literature on the Kutta condition in unsteady aerodynamics informs us that finite velocities at the trailing edge, or flow leaving smoothly from the trailing edge are necessary but insufficient conditions to correctly determine the unsteady circulation. An additional condition is required, which requires the pressure difference to be zero at the trailing edge, or vorticity to be non-zero and continuous across the trailing edge and into the wake. 

In the conventional implementation of unsteady thin-aerofoil theory, a discrete vortex is used to model the last shed vorticity at any given time, which results in vorticity at the trailing edge being zero. This violation of the second part of the Kutta condition in turn leads to non-zero values of pressure difference at the trailing edge which is not physically possible. To rectify this, a modified implementation of UTAT is introduced in this article, where the continuous wake vorticity distribution from Wagner's exact solution is used to represent the last shed vorticity at any time. The bound vorticity induced by the last shed wake and the wake vorticity itself are continuous across the trailing edge, satisfying the second part of the Kutta condition. The effects of this modification are illustrated for a variety of unsteady kinematic manoeuvres. In all cases, the new implementation was able to exactly recover the analytical and CFD "true" solutions with no errors, whereas the original implementation had an error associated with the Kutta condition being unsatisfied, that increased with increasing reduced frequency.

\section*{Acknowledgements}
This material is based upon work supported by the US Air Force Office of Scientific Research under award number FA8655-21-1-7018. The author thanks Prof. Sjoerd W. Rienstra (Eindhoven University of Technology) for sharing his insights on the Glauert series and the Kutta condition, which partially motivated this research.

\bibliographystyle{jfm}
\bibliography{kiran_bibtot}

\end{document}